# Protecting a solid-state spin from decoherence using dressed spin states

D. Andrew Golter, Thomas K. Baldwin, and Hailin Wang

Department of Physics, University of Oregon, Eugene, OR 97403, USA

Abstract

Dressed spin states, a spin coupling to continuous radiation fields, can fundamentally change how a spin responds to magnetic fluctuations. Using dressed spin states, we were able to protect an electron spin in diamond from decoherence. Dressing a spin with resonant microwaves at a coupling rate near 1 MHz leads to a 50 times reduction in the linewidth of the spin transition, limited by transit-time broadening. The spin decoherence and the energy level structure of the dressed states were probed with optical coherent-population-trapping processes. Compared with dynamical decoupling, where effects of the bath are averaged out at specific times, the dressed state provides a continuous protection from decoherence.

Complete quantum control of individual electron spins in solids provides exciting opportunities for quantum coherence-based applications[1]. Electron spins in solids, however, are susceptible to magnetic fluctuations in their surrounding environment. The electron spins couple to magnetic fields induced by nearby nuclear or electron spins. The fluctuations in this spin bath induce corresponding variations in the spin dynamics, shortening the spin coherence time. The protection of electron spins from this environment-induced decoherence has been a fundamental challenge in quantum science and technology. Previous experimental and theoretical efforts to overcome spin decoherence have focused on dynamical decoupling, using rapid spin flips to average out effects of magnetic fluctuations[2-11]. Dynamical decoupling approaches, such as the optimized Uhrig sequence or the periodic Carr-Purcell-Meibloom-Gill sequence, have been successfully implemented in diamond, semiconductors, and other spin systems[3-6]. With this time domain approach, an electron spin subject to the spin-flip pulse sequence is decoupled from the bath at specific times. The dynamics of the electron spin, however, are still influenced by the fluctuating magnetic field. In addition, the decoupling pulse sequence can often be in conflict with the desired quantum operations.

It has been suggested recently that an electron spin can be decoupled from the bath at all times with a spectral domain approach, in which a coherent coupling between the spin and continuous microwave fields leads to the formation of dressed spin states[12]. The energy levels of the dressed spin states can become immune to fluctuating magnetic fields, when the coherent coupling rate far exceeds the relevant amplitude and rate of the bath-induced fluctuations. In essence, the energy gap between the dressed spin states protects the electron spin from decoherence induced by the spin bath[12, 13].

Here we use an electron spin of a negatively-charged nitrogen vacancy (NV) center in diamond as a model solid-state spin system. We show that an electron spin in diamond can be protected from decoherence using the dressed spin states. Dressing a single electron spin with microwave fields at a modest coupling rate near 1 MHz leads to a 50 times reduction in the linewidth of the spin transition, limited by transit-time broadening. The dependence of the spin decoherence rate on the amplitude of the microwave dressing field further reveals that spin coherence arising from the same dressed states can be immune to magnetic fluctuations even when the energy gap of the dressed states is comparable to or even smaller than the relevant energy scale of the bath fluctuations. Dressed spin states fundamentally change how a spin responds to magnetic fluctuations, opening up a new area for using electron spins in applications such as quantum information processing and coupled spin-nanomechanical systems[14, 15].

To form dressed spin states, we couple two microwave fields resonantly to the two respective spin transitions of the $m_s=0$ and $\pm 1$ states of the ground-state spin triplet of a NV center, with equal Rabi frequency $\Omega_m$ (see Fig. 1a). The energy eigenstates in the rotating frame, i.e. the semiclassical dressed states, of this combined spin-microwave system are $|d>$, $|l>$, and $|u>$, where $|d>=(|+>-|->)/\sqrt{2}$ is a dark state decoupled from the microwave fields. The orthogonal bright state, $|b>=(|+>+|->)/\sqrt{2}$, couples to the microwave fields, leading to the formation of two other dressed states, $|l>=(|0>-|b>)/\sqrt{2}$ and $|u>=(|0>+|b>)/\sqrt{2}$. The eigen energies of the dressed spin states are $E_l = -\hbar\Omega_m/\sqrt{2}$, $E_d = 0$, and $E_u = \hbar\Omega_m/\sqrt{2}$.

The energy levels of states $|\pm>$ depend on the local magnetic field $\boldsymbol{B}_0+\delta\boldsymbol{B}$, where $\boldsymbol{B}_0$ is the fixed external magnetic field along a NV axis and leads to a Zeeman splitting of $\omega_B$ between the $|\pm>$ states. As illustrated in Fig. 1b, the bath-induced magnetic field, $\delta\boldsymbol{B}$, leads to an additional Zeeman shift, $\pm\delta_N$, for states $|\pm>$. Accounting for this bath-induced Zeeman shift, the eigen energies of the dressed states are then given by $E_l = -\hbar\sqrt{\Omega_m^2/2 + \delta_N^2}$ and $E_u = \hbar\sqrt{\Omega_m^2/2 + \delta_N^2}$, with $E_d$ remaining unchanged, as indicated in Fig. 1b[13]. In the limit that $\Omega_m >> |\delta_N|$, the energy levels of the dressed states become nearly independent of $\delta_N$. Furthermore, the dressed states enforce an energy and phase correlation between the $m_s=+1$ and $m_s=-1$ part of the same dressed states. The corresponding spin coherence can be immune to the magnetic fluctuations, even without $\Omega_m >> |\delta_N|$.

The microwave-induced dressed spin states can be probed through optical transitions. In a NV center, states $|\pm>$ couple to the $A_2$ excited state (denoted as $|e>$) via $\sigma_\mp$ circularly polarized optical fields[16, 17]. With $\Omega_m=0$, the $m_s=\pm 1$ and $A_2$ states form a nearly closed $\Lambda$-type three-level system, which has been used for spin-photon entanglement and also for optical control of electron spins[18, 19]. In the presence of the resonant microwave fields, the electron wave function can be described with probability amplitudes $C_d$, $C_l$, $C_u$, for the dressed states and $C_e$ for the excited state,

$$|\psi\rangle = (\frac{C_u}{2}e^{-i\Omega_m t/\sqrt{2}} + \frac{C_l}{2}e^{i\Omega_m t/\sqrt{2}} + \frac{C_d}{\sqrt{2}})e^{-i\omega_B t}|+\rangle + (\frac{C_u}{\sqrt{2}}e^{-i\Omega_m t/\sqrt{2}} - \frac{C_l}{\sqrt{2}}e^{i\Omega_m t/\sqrt{2}})e^{i\nu t}|0\rangle$$
$$(\frac{C_u}{2}e^{-i\Omega_m t/\sqrt{2}} + \frac{C_l}{2}e^{i\Omega_m t/\sqrt{2}} - \frac{C_d}{\sqrt{2}})|-\rangle + C_e|e\rangle$$
(1)

where $\nu$ is the frequency of the microwave field coupling to $|-\rangle$. Similar expressions can be derived for $\delta_N \neq 0$, with corresponding changes in the energy and wave function of the dressed states. As shown in Eq. 1, $|\pm\rangle$ each effectively splits into three different energy levels due to the coupling with the microwave fields. In this case, $\sigma_+$ and $\sigma_-$ polarized optical fields couple $|e\rangle$ to the $m_s = -1$ part ($|d_-\rangle$, $|l_-\rangle$, and $|u_-\rangle$) and $m_s = +1$ part ($|d_+\rangle$, $|l_+\rangle$, and $|u_+\rangle$) of the dressed spin states, respectively, as shown in Fig. 2a.

We have used coherent population trapping (CPT) to probe the energy level structure and decoherence rates of the dressed spin states[20-23]. For a Λ-type three-level system driven by two resonant optical fields with detuning $\delta$ and with equal Rabi frequency $\Omega_0$, the steady-state spin coherence between the two lower states with frequency separation $\omega_0$ is given by[13]:

$$\rho_{-+} = -\Omega_0^2(N_+ + N_-)(4\gamma)^{-1} / [i(\delta - \omega_0) + \gamma_s + \Omega_0^2/2\gamma] \quad (2)$$

where $\gamma_s$ and $\gamma$ (with $\gamma_s \ll \gamma$) are the decay rates for the spin and optical dipole coherence, respectively, $N_\pm$ is the population difference between the respective lower state and the excited state. With $\delta \approx \omega_0$ and $\Omega_0^2 \gg 2\gamma\gamma_s$, the CPT drives the system toward an optical dark state with $\rho_{-+} = -1/2$, preventing the optical excitation of the excited state[24].

Our experiments were performed in a type IIa diamond at 5 K. A 532 nm diode laser provided off-resonant optical excitations of the NV center and a 637 nm tunable ring laser provided resonant optical excitations. The frequency detuning between the two resonant optical fields was generated with acousto-optic modulators. The two microwave transitions were driven resonantly by outputs from two separate but phase-locked radio-frequency signal generators. The CPT experiments were carried out with resonant optical pulses with duration of 40 μs, alternated with off-resonant excitations in order to reverse any ionization of the negatively-charged NV center. We determined $\Omega_m$ by using Rabi oscillations driven by individual microwave fields and $\Omega_0$ by using Rabi oscillations driven by an optical Raman transition[19, 25]. An incident optical power of 1 nW corresponds to an estimated $\Omega_0/2\pi = 0.74$ MHz.

We obtained the CPT spectral response by measuring the emission from $|e\rangle$ as a function of the detuning between the two optical fields. For these studies, the NV was initially prepared in the $m_s=0$ state. Two microwave fields, with $\Omega_m$ comparable to the linewidth of the transitions between bare spin states, were resonant with the respective spin transitions. Two optical fields with opposite circular polarization and equal intensity were resonant with the respective $A_2$ transitions. The CPT spectral response shown in Fig. 2b features five resonances (or sharp dips), instead of a single resonance observed for the bare spin states[23]. Note that due to the hyperfine coupling between the electron spin and the $^{14}$N nuclear spin in a NV center, the energy of $|\pm\rangle$ also depends on the spin projection, $m_n$, of the $^{14}$N nucleus. Each electron spin state splits into three hyperfine states, with $m_n=-1, 0, +1$. The CPT resonances in Fig. 2b correspond to $m_n=0$. CPT resonances with $m_n=-1$ and $+1$ are 4.4 MHz away from that with $m_n=0$[21, 23].

For the energy level structure in Fig. 2a, spin coherences can occur between two different dressed states (for example, between $|l_+\rangle$ and $|u_-\rangle$) and can also arise from the same dressed states (for example, between $|d_+\rangle$ and $|d_-\rangle$). Spin coherences arising from the same dressed states lead to the central CPT resonance at $\delta=\omega_B$. Spin coherences between $|d\rangle$ and either $|l\rangle$ or $|u\rangle$ lead to CPT resonances at $\delta = \omega_B \pm \Omega_m/\sqrt{2}$, the first sidebands in Fig. 2b. Spin coherences between $|l\rangle$ and $|u\rangle$ lead to CPT resonances at $\delta = \omega_B \pm \sqrt{2}\Omega_m$, the second sidebands in Fig. 2b. Figure 2c plots the $\Omega_m$-dependence of the spectral separation between the first sidebands and the central resonance derived from experiments similar to that in Fig. 2b, confirming the above assignment.

For the central CPT resonance, all six spin states in Fig. 2a are involved in the CPT process, which can drive or pump the electron into an optical dark state. Since the relative phase of the two optical fields is typically different from that of the two microwave fields, the optical dark state can differ from $|d\rangle$, the microwave dark state. For an optical dark state of the form $|d_{opt}\rangle = (|+\rangle + e^{i\theta}|-\rangle)/\sqrt{2}$, the electron is in $|d_{opt}\rangle$ when $(C_l + C_u)/C_d = \sqrt{2}(1+e^{i\theta})/(1-e^{i\theta})$ (assuming $\Omega_m \ll \gamma$)[13]. For the first and second sidebands, only four and two spin states in Fig. 2a are involved in the CPT process, respectively (see the inset in Fig. 2b), leading to weaker CPT resonances. A detailed analysis shows that with $\Omega_0^2 \gg 2\gamma\gamma_s$, we expect a ratio of 100:70:25 for the amplitudes of the central resonance, first sidebands, and second sidebands of the CPT spectral response[13]. Figure 2b shows the good agreement between the experiment and the theoretical expectation, for which we used the above ratio, along with an overall scaling factor for the CPT amplitudes and a Lorentzian linewidth of 0.22 MHz.

The linewidth of CPT resonances is determined by the decay of the underlying spin coherence and also by power-dependent broadening mechanisms. For an ideal Λ-type system and at relatively low intensity, the effective linewidth is given by $2\gamma_s^{eff} = 2\gamma_s + \Omega_0^2/\gamma$ (see Eq. 2), scaling linearly with the optical power. We first discuss the behavior of the central CPT resonance. Figure 3a shows the linewidth of the central CPT resonance as a function of the input laser power, obtained with $\Omega_m/2\pi$=0.83 MHz and under otherwise similar conditions to Fig. 2b. The central resonance obtained at the lowest laser power used is shown in Fig. 3b. Figure 3c compares the CPT linewidth for the dressed spin states with that for the bare spin states obtained under otherwise similar conditions. For CPT of the bare spin states, the NV was initially prepared in the $m = +1$ state[23]. Figures 3a and 3c also show the theoretically calculated power-dependent CPT linewidth[13]. For these calculations, density matrix equations were used, with $\gamma/\pi$=13 MHz[26]. No adjustable parameters were used other than an overall scaling factor for the amplitude of the CPT resonances (this is needed since background contributions including dark counts and scattered laser light were not subtracted from the CPT spectral response). The calculations also included effects of NV spectral diffusion[13].

For the CPT linewdith of the bare spin states, deviation from linear power dependence occurs as $\Omega_0$ approaches $\gamma$. The deviation from linear power dependence for the dressed spin states, however, occurs at a much lower power, as shown in Fig. 3a. As $\Omega_0^2/\gamma$ approaches $\Omega_m$, spin coherences of the dressed spin states (or relevant Λ-type systems) are no longer independent of each other. The coupling between the spin coherences, which is included to the lowest order in our calculation[13], leads to a smaller CPT linewidth (i.e. smaller power-dependent broadening). The saturation in power broadening observed for the dressed spin states thus signals optically induced coupling between spin coherences of the dressed spin states.

From Fig. 3, we derive a spin transition linewidth without power-dependent broadening, $2\gamma_s/2\pi$ =0.75 MHz and 13 kHz, for the bare and dressed spin states, respectively. The large reduction in the spin transition linewidth demonstrates a spin coherence that is protected by the formation of the dressed states. The robustness of the spin coherence against magnetic fluctuations paradoxically also makes it difficult to measure the decoherence rate with a spectral domain technique. The intrinsic spin transition linewidth obtained with spin echoes in an isotopically pure diamond is 0.18 kHz[27]. Eliminating power broadening at this frequency scale requires diminishing incident laser powers. Another limitation is that resonant optical excitations of a negatively-charged NV center inevitably lead to electron ionization[28]. To avoid the ionization as well as undesired optical pumping effects[18], we used square optical pulses with duration of 40 μs. The resulting transit-time broadening sets a lower limit of about 12 kHz for the measured spin transition linewidth, in good agreement with the experiment.

The dependence of the CPT linewidth on $\Omega_m$ is shown in Fig. 3d. The first sideband and the central resonance feature similar power-broadened linewidths at relatively large $\Omega_m$. When $\Omega_m$ decreases below the linewidth of the bare spin transition, the linewidth of the first sideband becomes significantly greater than that of the central resonance. Note that although the bath-induced spin dephasing for the first sideband is suppressed to below 30 kHz at $\Omega_m=1$ MHz, the $\Omega_m$-dependence becomes observable only when the spin dephasing becomes significant compared with the power-dependent broadening[13]. Figure 3d demonstrates an important difference between spin coherences of different dressed states and spin coherences arising from the same dressed states. The protection of spin coherences of different dressed states requires that $\Omega_m$ be large compared with or at least comparable to the linewidth of bare spin transitions, which is not necessary for spin coherences arising from the same dressed state.

For quantum information processing, dressed spin states can be used directly as qubits. With $\Omega_m$ over 100 MHz, which has been achieved in earlier studies[29], a nearly complete suppression of spin dephasing can be attained. CPT of the dressed spin states discussed above also indicates the feasibility of performing quantum control of dressed spin qubits through off-resonant optical Raman transitions, similar to those realized for optical control of bare spin states[19, 22]. The dressed spin states can be especially useful when dynamical decoupling is in conflict with the desired quantum operations or when continuous coupling is required, such as the cooling of a mechanical oscillator via coupling to an electron spin[12]. With suitable microwave or optical transitions, coherence protection with dressed spin states can also be extended to other solid state spin systems such as SiC[30].

This work has been supported by NSF with grants No. 1005499 and No. 1104718 and by the DARPA-MTO ORCHID program through a grant from AFOSR.

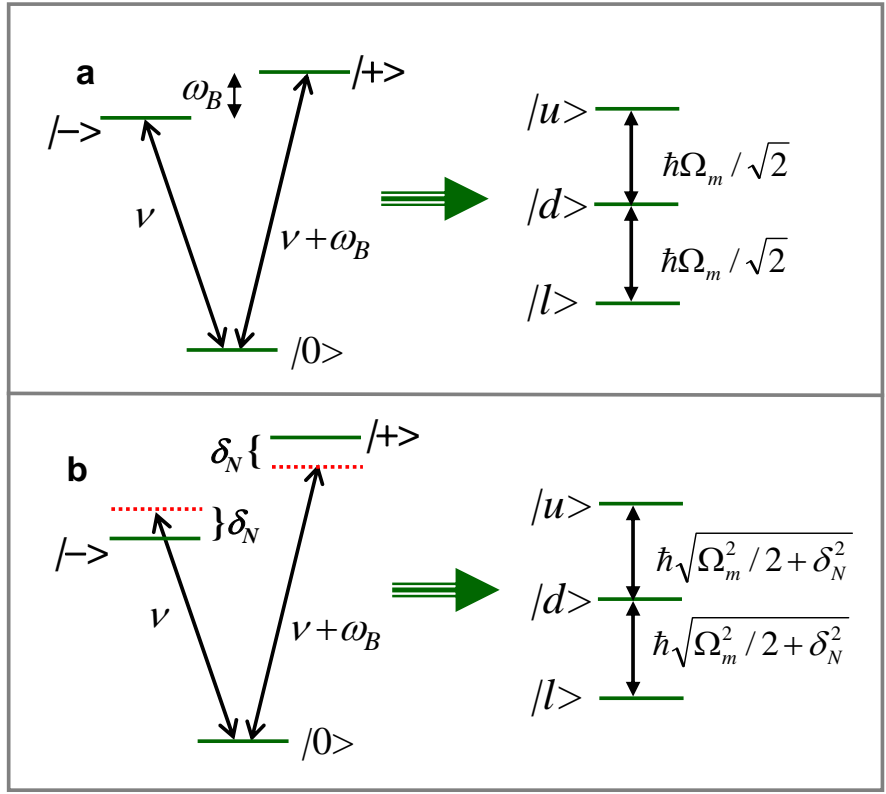

**Fig. 1** (a) The coupling of the NV ground-state spin triplet to two resonant microwave fields with equal Rabi frequency $\Omega_m$ leads to the formation of three dressed states, $|l\rangle, |d\rangle, |u\rangle$. (b) The energies of the dressed spin states, for which effects of bath-induced Zeeman shift, $\pm\delta_N$, are included. The dressed state energies become immune to the bath-induced magnetic fluctuations when $\Omega_m \gg |\delta_N|$.

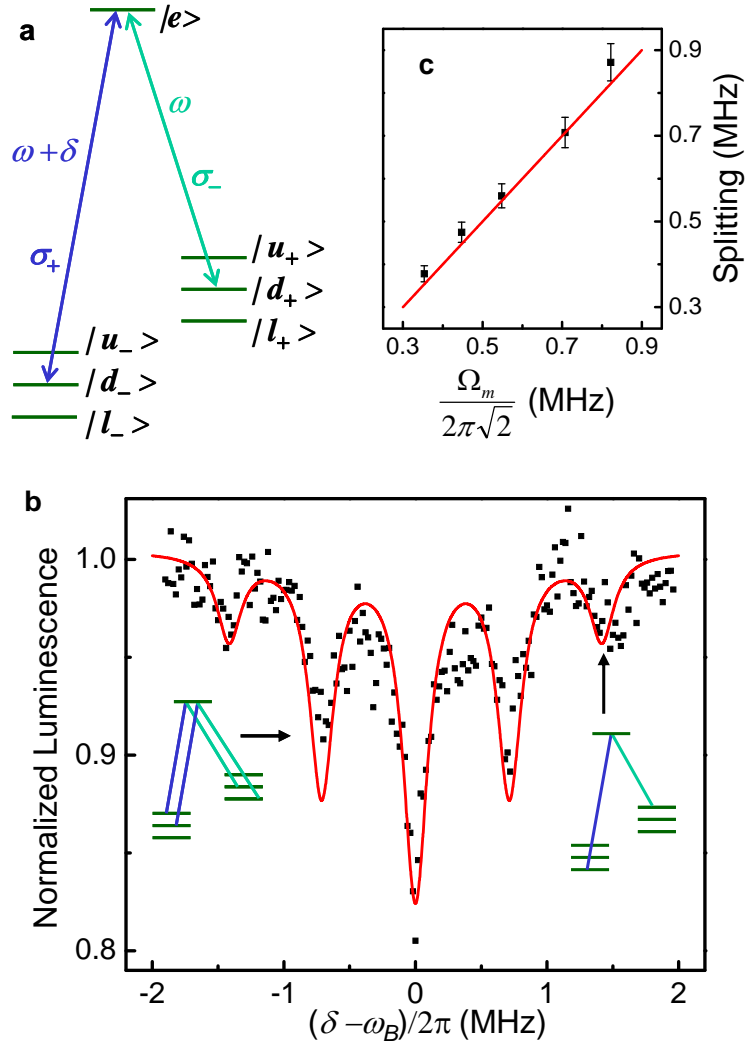

**Fig. 2** (a) Schematic of the $A_2$ excited state (denoted as |e>) coupling to the $m_s= -1$ part (|d_>, |l_>, and |u_>) and $m_s= +1$ part (|d_+>, |l_+>, and |u_+>) of the dressed spin states via $\sigma_+$ and $\sigma_-$ circularly polarized optical fields, respectively. (b) The CPT spectral response obtained with $\Omega_m/2\pi=1$ MHz and at an incident optical power of 6 nW. Spin coherences arising from the same dressed state lead to the central resonance. Spin coherences between |d> and either |l> or |u> and those between |l> and |u> lead to the first and second sidebands, respectively, as indicated in the inset. The red curve shows the result of a theoretical calculation discussed in the text. (c) The frequency splitting between the central resonance and the first sidebands as a function of $\Omega_m$. The solid line intercepts with the origin and has a slope of 1.

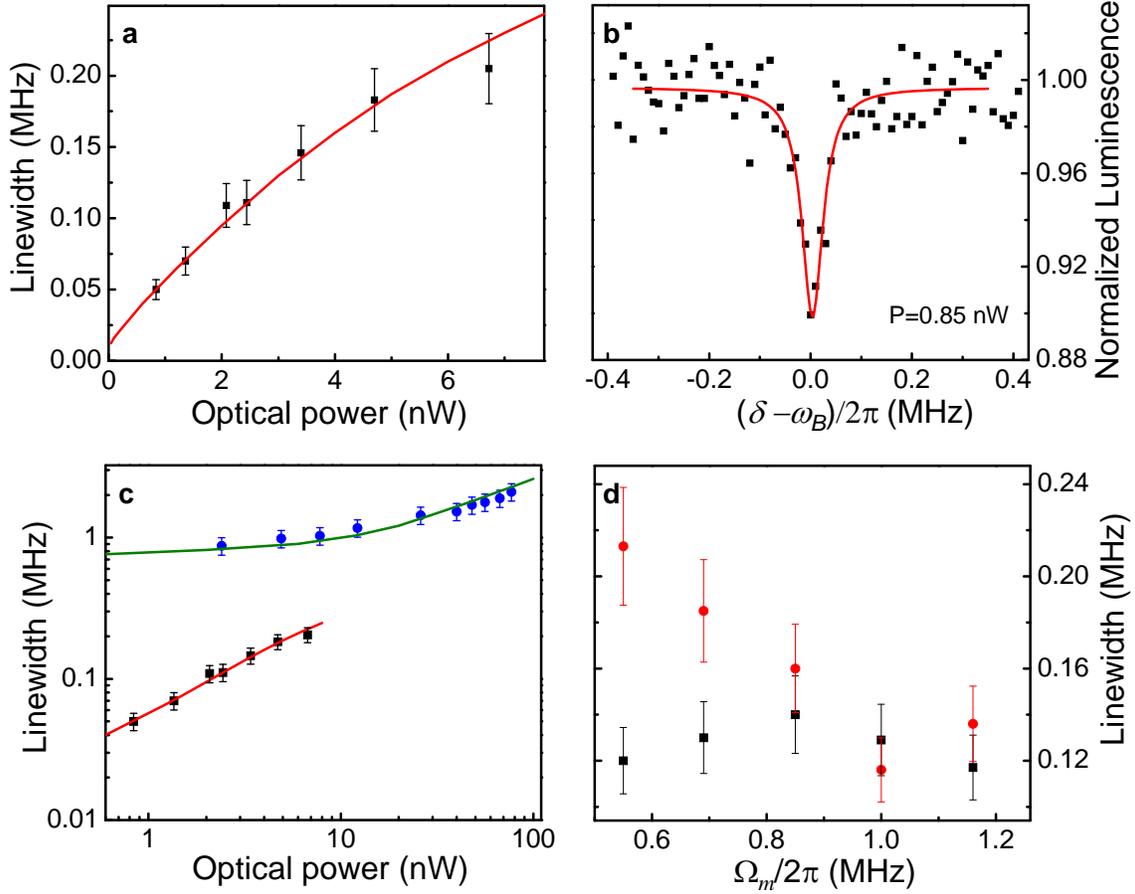

**Fig. 3** (a) The linewidth of the central CPT resonance for the dressed spin states as a function of the incident optical power, for which 1 nW corresponds to an estimated $\Omega_0/2\pi=0.74$ MHz. (b) The central CPT resonance obtained at the lowest optical power used. The solid line in (b) is a least square fit to Lorentzian. (c) Comparison between the linewidth of the CPT resonance for the bare spin states (dots) and that of the central CPT resonance for the dressed spin states (squares). Solid lines in (a) and (c) show the calculated power-dependent broadening. (d) The linewidth of the first CPT sideband (dots) and the central CPT resonance (squares) as a function of $\Omega_m$, obtained with an incident optical power of 2.5 nW.